\documentclass[pra,aps,twocolumn,showpacs]{revtex4}

\usepackage{amsmath,amsfonts,graphicx,epstopdf}
\usepackage{epsfig}
\usepackage{amssymb}
\usepackage{bm}
\usepackage{psfrag}

\newcommand{\ra}{\rangle}
\newcommand{\la}{\langle}

\begin{document}
\author{Rachele Fermani}
\email{rachele.fermani@googlemail.com} 
\author{Stefan Scheel}
\title{Screening of electromagnetic field fluctuations by $s$--wave and
$d$--wave superconductors} 
\affiliation{Quantum Optics and Laser Science, Blackett Laboratory,
Imperial College London, Prince Consort Road, London SW7 2BW}
\begin{abstract}
We investigate theoretically the shielding of the electromagnetic
field fluctuations by $s$--wave and $d$--wave superconductors within
the framework of macroscopic quantum electrodynamics. The spin flip
lifetime is evaluated above a niobium and a bismuth strontium calcium
copper oxide (BSCCO) surface, and the screening effect is studied as a
function of the thickness of the superconducting layer. Further, we
study the different temperature dependence of the atomic spin
relaxation above the two superconductors.  
\end{abstract}
\date{\today}
\maketitle

\section{Introduction}
Recent advances in the trapping of neutral atoms near microstructured
surfaces \cite{specialissue,fortagh}, colloquially referred to as atom
chips, has proved these structures to be promising for the coherent
manipulation of cold atom clouds. Applications of atom chips include
the imaging of the electromagnetic field near metallic and dielectric
surfaces \cite{Schmiedmayer05,Hall06}, the accurate measurement of
gravity \cite{clade,gravity} and atom interferometry
\cite{schumm,gunther}. One of the central challenges proves to be the
interaction between the trapped atoms and the hot surfaces in their
vicinity, which leads to thermally-induced spin relaxation dynamics
and attractive Casimir--Polder forces. The atomic lifetime is limited
by the transitions between trapped and untrapped states induced by
fluctuating electromagnetic fields arising in conducting materials
\cite{Jones03,spinflip}.    

The origin of the electromagnetic field fluctuations lies in the
finite resistivity (and the finite temperature) of conducting
materials as a consequence of the fluctuation-dissipation
theorem. Superconductors have a (almost) negligible resistivity and
their thermal fluctuations are attenuated with respect to a dielectric
or a metal. 
An important enhancement of the atomic lifetime is expected with the
adoption of superconducting surfaces, and both theoretical and
experimental investigations have recently turned their attention to
superconducting materials
\cite{Scheel05,Haroche06,Shimizu07,Scheel07,dumke,folman,cano}.

In this article, we investigate the shielding of the near-field noise
in the vicinity of a dielectric substrate provided by two different
superconducting films. Casimir--Polder potential and thermally induced
spin flips are among the effects which may be altered when a
superconducting layer is placed above a dielectric structure, and we
will focus on the latter.  

Previous theoretical investigations on superconducting atom chips have
been concerned with the increase of the spin flip lifetime and vortex detection via spin dynamics
\cite{Scheel05,Scheel07}. Three different descriptions of superconductivity
have been adopted to estimate the spin flip rate \cite{Ulrich07}: the
two-fluid model, the Bardeen--Cooper--Schrieffer (BCS) theory and the
Eliashberg theory. The Eliashberg theory provides the most appropriate
model for the electromagnetic energy dissipation in superconducting
materials, whereas the other two theories ignore the strong
modification of either the imaginary part or the real part of the
optical conductivity in the superconducting state.  Nevertheless, the
spin flip lifetime evaluated with the two-fluid model gives
remarkably accurate estimates of the Eliashberg results
\cite{Ulrich07}. 

Quantum mechanical approaches for spin flip lifetime calculations above superconducting atom chips have considered
only isotropic conventional ($s$-wave) superconductors.  
In this paper we introduce a comparison between the screening
properties of a conventional $s$-wave superconductor such as niobium,
and an unconventional $d$-wave superconductor such as bismuth
strontium calcium copper oxide (BSCCO).  
The main difference between these two classes of materials is the way
in which the electrons interact with each other to form Cooper pairs
\cite{Kleiner}.   
In conventional $s$-wave superconductors, the two electrons of a
Cooper pair are in a state with zero total spin and total angular
momentum, such that their state is isotropic and the superconducting
phase is homogeneous with an uniform penetration depth $\lambda$.

In $d$-wave superconductors, an unconventional pairing takes place and
the state of the paired electrons has a zero total spin while the
total angular momentum is $2 \hbar$ (hence the name). Such
superconducting state is anisotropic giving rise to layered
superconductors, which are characterised by an anisotropic
permittivity. This results from the in-plane penetration depth 
$\lambda_{\parallel}$ being different from the longitudinal
penetration depth $\lambda_{\perp}$. 

As an example, we report the values of the penetration depth for the
two superconducting materials considered in this work. For zero
temperature, niobium has a penetration depth of $\lambda(0)=35$~nm
\cite{miller}, whereas the transverse and in plane
penetration depth of BSCCO are $\lambda_{\perp} (0)=100\, \mu$m and
$\lambda_{\parallel}(0)=300$ nm, respectively \cite{Kleiner94}.  

Furthermore, $d$-wave superconductors have rather high transition
temperatures. For example, for BSCCO one finds transition temperatures
of up to $T_C=90$~K while for thin niobium films one has
$T_C=8.3$~K (compared to $T_C=9.3$~K for bulk niobium). The origin
lies in the crystal structure of unconventional superconductors \cite{Kleiner}.
The interlayers act as a reservoir of charge carriers that combine
into Cooper pairs and, depending on the concentrations of oxygen atoms
into the material, the transition temperature increases from small
values up to a maximum value of 90 K.  

Not only the value of the transition temperature, but also the
temperature dependence of the penetration depth differs in both
types. In conventional superconductors, the deviation of the
penetration depth $\lambda$ from its value at $T=0$ can be
approximated by the empirical formula \cite{Kleiner} 
\begin{equation}\label{eq:penS}
\frac{ \lambda(T)}{\lambda(0)} =  \left [ 1- \left (
\frac{T}{T_{C}}\right )^4\right ]^{-1/2}\, , 
\end{equation} 
while in $d$-wave superconductors, $\lambda_{\parallel}(T) /
\lambda_{\parallel}(0)$ increases with $T$ as
\cite{Kleiner,Bonalde}
\begin{equation}\label{eq:penD}
\frac{ \lambda_{\parallel}(T)}{\lambda_{\parallel}(0)} =
\left [ 1- \frac{T}{T_{C}}\right ]^{-1/2}. 
\end{equation} 

In present experiments on superconducting atom chips, the
superconductor is deposited on an insulator, as its contact with a
metal can perturb the superconducting properties. However, in order to
emphasize the screening properties of a superconducting medium, we
consider the superconducting film to be ideally placed above a metal
substrate. We emphasize
here that all the results obtained are valid only for the
superconductors in the Meissner state. The Meissner state is observed
when a magnetic field in any point of a superconductor is below the
first critical field, which is 140 mT for niobium and 13 mT for BSCC0
at 4.2 K. In an higher field, the magnetic field penetrates
into the superconductor (type-II) in the form of vortices. In this
so-called  mixed state a superconductor exhibits different screening
properties and dependences on frequency and temperature from the
Meissner state, hence the results presented here are not valid for the
mixed state.

The paper is organized as follows. In Sec.~\ref{sec:fluctuations}, we
give a brief introduction into vacuum fluctuation effects with emphasis
on spin flip transitions and the Lamb shift. The expressions for the
spin flip rate in isotropic and anisotropic planar multilayered
structures is discussed in detail in Sec.~\ref{sec:rates} and in the
Appendix. The results on atomic spin relaxation are given in
Sec.~\ref{sec:middle} with the screening properties investigated in
Sec.~\ref{sec:screening} and the temperature dependence of the
penetration depth investigated in Sec.~\ref{sec:temperature}. Our
conclusions are given in Sec.~ \ref{sec:conclusion}. 

\section{Vacuum fluctuations}
\label{sec:fluctuations}

The spontaneous decay of an excited atom is one of the most widely
studied effects of the ground-state fluctuations of the
electromagnetic vacuum \cite{MILONNI}. However, the presence of a
dielectric or metallic body strongly modifies the vacuum field and its
statistical properties \cite{Acta}, leading to a modification of the
spontaneous emission rate as well as the Lamb shift. In particular,
the latter is associated with the appearance of an attractive
dispersion force, the Casimir--Polder force.
The energy shift is position-dependent and takes the form of
$\Delta E = \Delta^{(0)} E + U(\mathbf{r}_A)$,
where the first term $\Delta^{(0)} E$ is the contribution to the Lamb
shift in free space, while the second term $U(\mathbf{r}_A)$ is the
position-dependent van der Waals potential of an atom at position
$\mathbf{r}_A$.
Another
related
issue associated to the modification of the vacuum
field is the body-induced modification of the atomic spin flip
lifetime \cite{spinflip}, which is the main subject of this paper.

Due to the Meissner effect, the presence of a superconducting material
provide a shield for the field fluctuations arising from a conducting substrate.  
The energy gap of a superconductor is
$\Delta_{g}=\hbar \omega_{g} \approx k_{B} T_{C}$, with $k_B$ the
Boltzmann constant, and corresponds to the minimum excitation energy
required to break a Cooper pair. In $s$-wave superconductors, the gap
frequency is almost of the order of THz (e.g. $\omega_{g} \approx 700$ GHz
for niobium \cite{bandgapN}), while in $d$-wave superconductors it is
usually an order of magnitude larger due to the higher transition
temperature (e.g. the BSCCO gap frequency is given in the literature
as 7.5 THz \cite{bandgapB}). 
All the physical phenomena resulting from field fluctuations are going
to be attenuated as long as the range of frequencies involved are smaller
than the gap frequency.

The atomic spin flip lifetime is enhanced by the presence of a superconductor as the typical transition
frequencies are of the order of MHz and fall well below the gap frequency
for both $s$-wave and $d$-wave superconductors. The same cannot be
said straightaway for the van der Waals potential. The induced energy
shift and broadening results in the atom having a position-dependent
polarizability associated with the relevant dipole transitions of the
atom. The potential $U(\mathbf{r}_A)$ is given by the sum of two terms: a contribution resonant with the
magnetic dipole transitions $U^{r}(\mathbf{r}_A)$ and an off-resonant
contribution $U^{or}(\mathbf{r}_A)$ \cite{Buhmann_C}. For an atom in
its ground state, the resonant term vanishes and the off-resonant term
is obtained by integrating over all the frequencies, which means that an
attenuation of the potential for frequencies $\omega < \omega_g$ is
not relevant on the overall effect. Hence, the Casimir--Polder force
experienced by a ground state atom is largely unaffected
by the presence of either $s$-wave or $d$-wave superconductors.  
However, the resonant term $U^{r}(\mathbf{r}_A)$ may dominate
$U^{or}(\mathbf{r}_A)$ for an atom in an excited state and a
superconductor may play a
significant role. Typical optical transition frequencies for an atom are of the
order of THz, and $d$-wave superconductors in principle may be able to alter their
Casimir--Polder potential.

\subsection{Spin flip rate}
\label{sec:rates}

\begin{figure}[h!]
\begin{center}
\includegraphics[width=8.5cm]{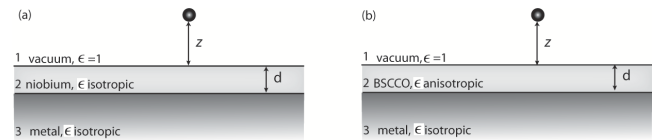}
\end{center}
\vspace{-0.5cm}
\caption{Schematic representation of the geometry of the two
structures considered: a three-layered geometry consisting of (a)
vacuum-niobium-metal  or (b) vacumm-BSCCO-metal. }\label{fig:layerS}
\end{figure}

The spin flip rate of an atom placed near a conducting or
superconducting body is obtained within the formalism of macroscopic 
quantum electrodynamics as \cite{spinflip} 
\begin{eqnarray}
\label{eq:sp}
\Gamma
&=& \mu_{0} \frac{2 (\mu_{B} g_{S})^2}{\hbar} 
\\
&& \times 
\la f | \hat{S}_{j}|i \ra 
\la i | \hat{S}_{k}|f \ra \,
\mathrm{Im} [\overrightarrow{\bm{\nabla}} \times \bm{G}(\mathbf{r}_A,\mathbf{r}_A,\omega) \times \overleftarrow{\bm{\nabla}} ]_{jk} \, ,
\nonumber
\end{eqnarray}
where $\la f | \hat{S}|i \ra$ is the spin matrix element for the
relevant transition, $\mu_{B}$ denotes the  Bohr magneton, and $g_{S}
\approx 2$ is the electron's $g$ factor. The Green function
$\bm{G}(\mathbf{r},\mathbf{r}',\omega)$ is the fundamental solution to
the classical Helmholtz equation and thus contains all necessary
geometric and electromagnetic information about the macroscopic
bodies. We review in the Appendix how to evaluate the Green tensor for
an anisotropic multilayered planar structure such as the one represented
in Fig.~\ref{fig:layerS} consisting of vacuum, superconducting layer
and metal substrate.

The spin flip rate for a planar multilayered structure made of
isotropic layers is known to be \cite{Fermani06}
\begin{equation}\label{eq:spinRateOLD}
\Gamma_{s} = \mu_0 \frac{ (\mu_{B} g_{S})^2}{8  \hbar } \!\! \int
\frac{K^2 dK}{(2 \pi)^2} \frac{e^{-2 K z}}{2}
\mathrm{Im}[\tilde{r}_{12}^{\scriptscriptstyle{TE}}], 
\end{equation}
where the spin matrix elements have been evaluated for the $^{87}$Rb
ground state transition between the magnetic hyperfine sublevels
$|2,2 \rangle \rightarrow |2,1 \rangle$. Let us recall here that
Eq.~(\ref{eq:spinRateOLD}) is valid whenever the atomic transition
wavelength can be regarded as the largest wavelength of the system.  
The generalized Fresnel reflection coefficient
$\tilde{r}_{12}^{\scriptscriptstyle{TE}}$ for a three-layer geometry
reads 
\begin{equation}
\tilde{r}_{12}^{\scriptscriptstyle{TE}}
=\frac{r_{12}^{\scriptscriptstyle{TE}}+r_{23}^{\scriptscriptstyle{TE}}
e^{2 i k_{2z} d}}{1- r_{21}^{TE} r_{23}^{\scriptscriptstyle{TE}} e^{2
i k_{2z}d }}\, , 
\end{equation}
where $d$ is the thickness of layer 2. The function
$r_{12}^{\scriptscriptstyle{TE}}$ is the Fresnel reflection
coefficient for TE waves at a planar interface defined as 
\begin{equation}\label{eq:r12}
r_{12}^{\scriptscriptstyle{TE}}=\frac{ k_{1z}-k_{2z} }{ k_{1z}+k_{2z} }, 
\end{equation}
with $k_{iz}^2=k_{i}^2-K^2$ and
$k_{i}^2=(\omega^2/c^2)\epsilon_{i}(\omega)$ with the label $i$
indicating the layer. The coefficient
$r_{23}^{\scriptscriptstyle{TE}}$ has the same form 
as $r_{12}^{\scriptscriptstyle{TE}}$ after the replacements
$k_{1z} \rightarrow k_{2z}$ and $k_{2z} \rightarrow k_{3z}$ have been
made.

The dielectric function $\epsilon(\omega)$ for a normal metal is
given by $\epsilon(\omega)\approx 2i\epsilon_0/ k^2 \delta_m^2$
where $\delta_m$ is the skin depth of the metal and $k=\omega/c$.  
In the two-fluid model \cite{Kleiner}, the dielectric function
$\epsilon(\omega)$ of a superconductor can be written as 
\begin{equation}\label{eq:perm}
\epsilon (\omega) = 1- \frac{1}{k^{2} \lambda^{2}(T)} + i
\frac{2}{k^{2} \delta^{2}(T)}\, ,
\end{equation}
where $\delta(T)= \sqrt{2/ \omega \mu_{0} \sigma_{n}(T)}$ and
$\sigma_{n}(T)$ are the skin depth and  the conductivity associated
with the normally conducting electrons, respectively. The optical
conductivity corresponding to Eq.~(\ref{eq:perm}) is 
\begin{equation}\label{eq:cond}
\sigma (\omega) = \frac{2}{\omega \mu_{0} \delta^{2}(T)}+
\frac{i}{\omega \mu_{0} \lambda^{2}(T)}. 
\end{equation}
Both the penetration depth and the skin depth are temperature
dependent \cite{Kleiner}, 
\begin{eqnarray}\label{eq:lambda}
\lambda(T)= \frac{\lambda(0)}{\sqrt{n_{s}(T)/n_{0}}}\, , \quad
\sigma_{n}(T)= \sigma \frac{n_{n}(T)}{n_{0}}\, , 
\end{eqnarray}
where $\sigma$ is the electrical conductivity of the metal in the
normally conducting state for $T>T_C$, and $n_{s}(T)$ and $n_{n}(T)$
are the electron densities of the superconducting and normal state,
respectively, at a temperature $T < T_{C}$. The total electron density
$n_{0}=n_{s}(T)+n_{n}(T)$ is constant, with $n_{s}(T)=n_{0}$ for $T=0$
and $n_{n}(T)=n_{0}$ for $T \geq T_{C}$. It follows from
Eq.~(\ref{eq:penS}) that Eq.~(\ref{eq:lambda}) can be written as
\begin{eqnarray}
\frac{\lambda (T)}{\lambda(0)} = \left [ 1-
\frac{n_{n}(T)}{n_{0}}\right ]^{-1/2} \!\!\!\!\!\! 
=\left [ 1- \left ( \frac{T}{T_{C}}\right )^\alpha \right ]^{-1/2}\!\!\!\!,
\end{eqnarray}
with $\alpha=4$ for an $s$-wave superconductor, and $\alpha=1$ for a
$d$-wave superconductor.

In contrast, in a $d$-wave superconductor the penetration depth and
thus the permittivity is highly anisotropic so that it can be written
as a tensor $\overline{\epsilon}$ of the form 
\begin{equation}\label{eq.permittivity}
\overline{\epsilon}= 
\left (
\begin{array}{ccc}
\epsilon_{t} & 0 & 0 \\
0 & \epsilon_{t} & 0 \\
0 & 0 & \epsilon_{z}
\end{array}
\right ) = \epsilon_{t} (\hat{\mathbf{x}} \hat{\mathbf{x}}
+\hat{\mathbf{y}} \hat{\mathbf{y}})  + \epsilon_{z} \hat{\mathbf{z}}
\hat{\mathbf{z}}\, , 
\end{equation}
where $\epsilon_{t}$ and $\epsilon_{z}$ are the transverse and
longitudinal scalar permittivities of the layer corresponding to
$\lambda_{\parallel}$ and $\lambda_{\perp}$, respectively.
We calculate the spin flip rate for the BSCCO structure as  
\begin{eqnarray}\label{eq:spinRateD}
\Gamma_{d}\! = \! \mu_0 \frac{ (\mu_{B} g_{S})^2}{8  \hbar }  
\!\!\int\limits_{0}^{\infty} d \eta \, \frac{ e^{-2  \eta z}}{8 \pi}
\mathrm{Im}\!\left [ 3 \eta^2 B_{M 1}^{11}
+ B_{N 2}^{11} k_{1}^{2}
\right ] ,
\end{eqnarray}
where $B_{M 1}^{11}$ and $B_{N 2}^{11}$ are the relevant scattering
coefficients given in the Appendix. 
We assume that the whole structure of metallic substrate plus
superconductor is in thermal equilibrium with its surroundings. The
electromagnetic field is then in a thermal state with temperature $T$,
equal to the temperature of the materials. Therefore, both spin flip
rates in Eq.~(\ref{eq:spinRateOLD}) and Eq.~(\ref{eq:spinRateD}) need
to be multiplied by a factor $(n_{\mathrm{th}}+1)$ where the mean
thermal photon number is
\begin{equation}
n_{\mathrm{th}}=\frac{1}{e^{\hbar \omega / k_{B} T}-1}\,.
\end{equation}

\section{Atomic spin relaxation}
\label{sec:middle}

In this section, we study the atomic spin relaxation above a niobium
and a BSCCO surface. We imagine that a $^{87}$Rb atom is held at a
distance $z$ from the superconducting surface as shown in
Fig.~\ref{fig:layerS}. The spin flip lifetime $\tau=1/\Gamma$ is
evaluated for the $^{87}$Rb ground-state transition
$|2,2\rangle\rightarrow |2,1\rangle$ with the transition frequency
taken to be $f=560$ kHz.  
The comparison between niobium and BSCCO is done in terms of the
screening of the electromagnetic field fluctuations affecting the
atomic spin dynamics. Further comparison is done by studying the
temperature dependence of the spin flip lifetime.

\subsection{Screening}
\label{sec:screening}

\begin{figure}[h!]
\begin{center}
\includegraphics[width=8.5cm]{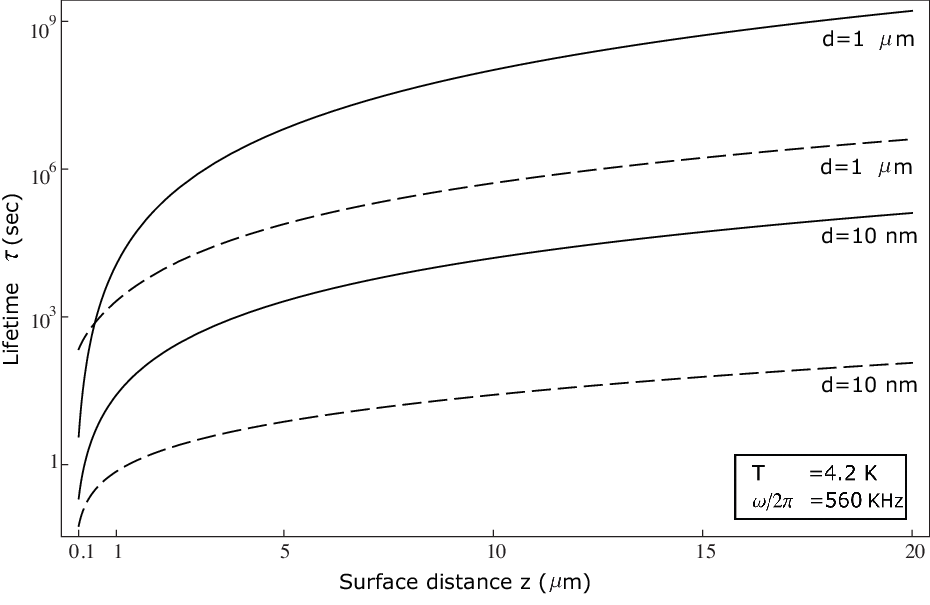}
\end{center}
\vspace{-0.5cm}
\caption{Atomic spin-flip lifetime near a niobium and a BSCCO
surface plotted as a function of the distance from the surface $z$ at
$T=4.2$ K.  Continuos line: lifetime above niobium. Dashed line:
lifetime above BSCCO. Two different thicknesses $d$ of the
superconducting layer are considered: $d=10$~nm and $d=1 \,
\mu$m.}\label{fig:SandD}
\end{figure}

The alteration of the spin flip lifetime as a function of the relevant
length scales of the system is an indication of the capabilities of
niobium and BSCCO to screen the electromagnetic field fluctuations
that arise from the underlying metal substrate. We consider the
multilayered structures as represented in Fig.~\ref{fig:layerS} and we
plot the atomic spin flip lifetime as a function of the distance $z$
from the surface in Fig.~\ref{fig:SandD}.  Two different thicknesses
of the superconducting layer are taken into consideration. 

For both niobium and BSCCO, the lifetime increases with the thickness
of the superconducting layer, and the longest lifetimes are obtained
for the atom trapped above niobium (continuous line in
Fig.~\ref{fig:SandD}).  
During the superconducting phase, magnetic field lines are expelled
from the material except for a thin surface layer (with thickness of
the order of the penetration depth), where shielding currents
flow. Therefore, fluctuating magnetic fields are attenuated by the
shielding currents. 
We now define a screening factor as a function of the thickness $d$ of
the superconducting layer as  
\begin{equation}
S(d)=\frac{\tau(d)-\tau(d=0)}{\tau(d=0)},
\end{equation}
where $\tau (d=0)$ is the spin flip lifetime associated with the bare
copper substrate obtained by combining Eq.~(\ref{eq:spinRateOLD}) and
Eq.~(\ref{eq:r12}). 

\begin{figure}[h!]
\begin{center}
\includegraphics[width=8cm]{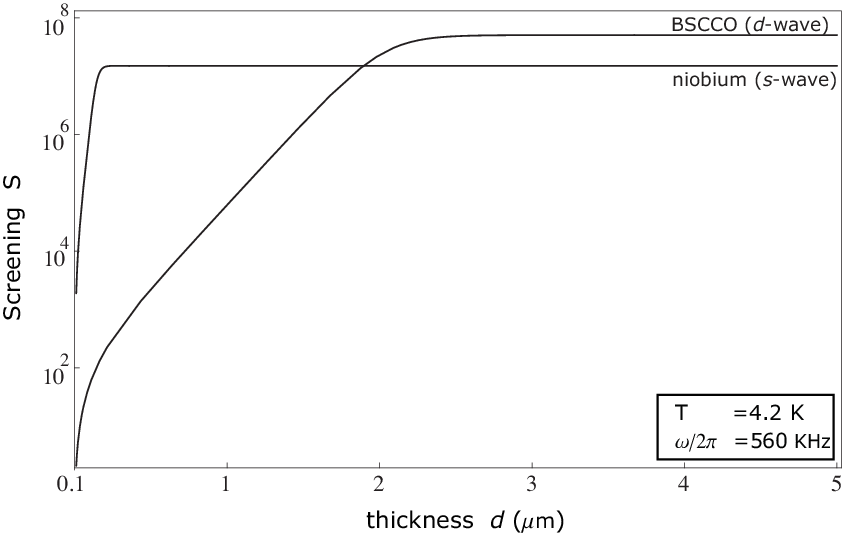}
\end{center}
\vspace{-0.5cm}
\caption{Screening factor $S(d)$ as a function of the thickness $d$
for an atom held at a distance of $10\, \mu$m from the superconducting
surface.} 
\label{fig:screen}
\end{figure}

This screening factor $S(d)$ is plotted in Fig.~\ref{fig:screen} for
niobium and BSCCO films and for an atom held at a fixed $10\,\mu$m
distance from the surface. The screening $S(d)$ increases in both
structures for thicknesses $d$ roughly the same size as the
penetration depths at zero temperature (recall that $\lambda(0)=35$~nm
for niobium and $\lambda_{\|}(0)=0.3\, \mu$m for BSCCO). For a
superconducting layer that is thicker that the corresponding 
penetration depth, the screening effect saturates which confirms the
fact that only the penetration depth provides an active screening of
the electromagnetic field fluctuations as already seen for dielectrics
\cite{Scheel05,Fermani06}.  

We emphasize
that the results presented in this paragraph are only indicative
for the two superconducting structures. The superconducting properties
change both with the layer thickness being less than the coherence
length (i.e.~30 nm for niobium at 4K). 

\subsection{Temperature dependence}
\label{sec:temperature}

Let us now investigate the temperature dependence of the spin flip
lifetime.
We have chosen a different thickness for each of the superconductors:
$1\,\mu$m  for niobium and $2.5\,\mu$m for BSSCO, corresponding to
the thickness providing the maximum screening (see Fig.~\ref{fig:screen}).
As shown in Eq.~(\ref{eq:penS}) and Eq.~(\ref{eq:penD}), the penetration
depth $\lambda$ is a function of the temperature, which is reflected in
the temperature dependence of the optical conductivity in
Eq.~(\ref{eq:cond}). 
However, the spin flip lifetime is proportional to the mean thermal photon
number $n_{\mathrm{th}}$, $\tau \propto 1/ (n_{\mathrm{th}}+1)$, which
may become dominant for large enough temperatures.  
 
\begin{figure}[h!]
\begin{center}
\includegraphics[width=8.5cm]{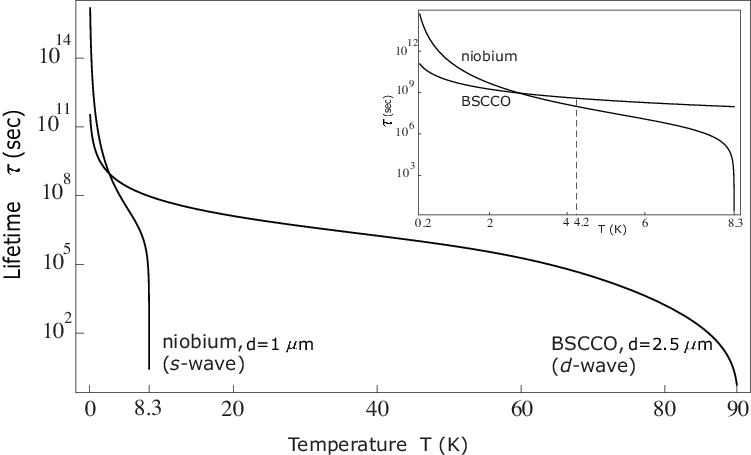}
\end{center}
\vspace{-0.5cm}
\caption{Spin-flip lifetime for the niobium and the BSCCO as a
function of the temperature $T$. In the inset, the spin-flip lifetimes
are plotted on a smaller temperature range $0.2-8.3$ K where $T=4.2$ K
refers to the liquid helium temperature. The atom-surface distance is
fixed at $z= 10\, \mu$m.   }\label{fig:tempFull} 
\end{figure}

\begin{figure}[h!]
\begin{center}
\includegraphics[width=8.5cm]{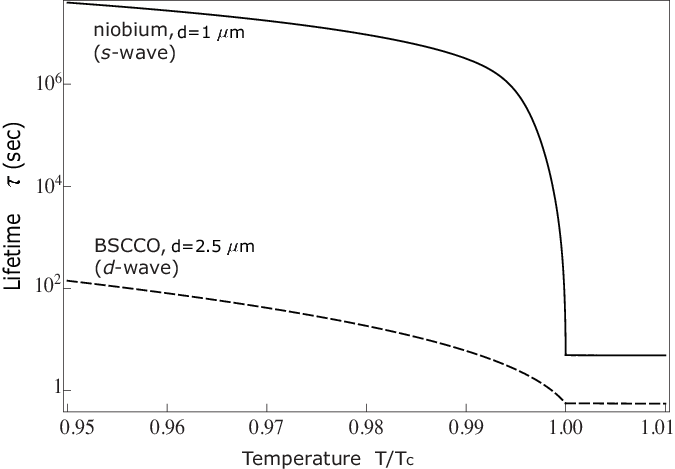}
\end{center}
\vspace{-0.5cm}
\caption{Spin flip lifetime for the niobium and BSCCO as a function of
$T/T_{C}$. The range chosen in around $T/T_{C}= 1$ in order to
highlight the power law of the lifetime at the transition
temperature. The atom-surface distance is fixed at $z= 10\,
\mu$m. }\label{fig:tempFullTC} 
\end{figure}

The comparison is carried out in two stages. First, in
Fig.~\ref{fig:tempFull} we show the lifetime $\tau$ as a function of
temperature, while in Fig.~\ref{fig:tempFullTC} we show $\tau$ as a
function of $T/T_{C}$, in order to study how the different power laws
affect the lifetimes near the transition temperature.
The inset in Fig.~\ref{fig:tempFull} shows the spin flip lifetime for
temperatures $T< T_{C_{s}}$ where both niobium and BSCCO are
superconducting. The lifetime near superconducting niobium
is several orders of magnitude larger than near superconducting
BSSCO. In particular, at the liquid helium temperature $T=4.2$~K, the
two lifetimes are approximately $\tau_{s}= 10^{10}$~s and $\tau_{d} =
5\,10^{6}$~s as shown in the inset of Fig.~\ref{fig:tempFull}. At the
liquid nitrogen temperature $T=77$ K, only BSCCO is still in the
superconducting phase giving a lifetime of roughly $\tau_{d}= 95$ s,
while the lifetime above (normally conducting) niobium is estimated to
be $\tau_{s} \approx 10^{-2}$ s. 

In Fig.~\ref{fig:tempFullTC}, the spin flip lifetime is plotted as a
function of $T/T_{C}$ based on Eqs.~(\ref{eq:spinRateOLD}) and
(\ref{eq:spinRateD}). For temperatures $T/T_{C}>1$, we neglect the
$\lambda$ dependence of the permittivity in Eq.~(\ref{eq:perm}) as
$n_{s}(T)=0$. For $T/T_{C} \rightarrow 1$, the lifetime above BSCCO
decreases with a smaller power law compared to niobium as a results of
the penetration depth dependence on temperature. 
There is a marked difference in the transition from the normal state
to the superconducting state, and we expect this to be visible in
experiments comparing the two structures considered here. According to
our results, the temperature dependence of the permittivity in
$d$-wave and $s$-wave superconductor can then be tested by observing
the atomic relaxation rate near the transition temperature.  

We would like to draw the attention to the orientation of the atomic
spin and how the spin flip rate changes according to it. The results
presented in this section have been obtained for a randomly oriented
spin. Due to the anisotropic permittivity of BSCCO, the shielding
of the electromagnetic field fluctuations can in principle be
different depending on which component couples to the atomic
spin. However, our estimates of the spin flip lifetime for a spin
oriented either perpendicular or parallel to the interface suggest
that the difference is not very pronounced. In fact, the spin flip
lifetimes for the two orientations differ only for a factor of 2, the
parallel orientation giving the longer lifetime, and it does not
depend on the anisotropic permittivity. This indicates that, despite
the transverse penetration depth being of the same order of magnitude
as in metals, electromagnetic field fluctuations are shielded by
screening currents independently from the anisotropy. 

\section{Conclusion}
\label{sec:conclusion}
We have considered the shielding of the electromagnetic field
fluctuations provided by a superconductor. The modification of the
vacuum field and its properties in the vicinity of a dielectric leads
to the appearance of the Casimir--Polder potential and
thermally-induced spin flip transitions. The frequencies associated
with these two phenomena are very different and only when they fall
below the superconducting gap frequency, the presence of the
superconducting material is detectable. While the frequencies
associated with spin transitions are always smaller than the gap
frequency, the same can not be said for the frequencies involved in
the Casimir--Polder potential and numerical calculations need to be
done.  

In this paper we have focused on the atomic spin relaxation above a
niobium and a BSCCO surface. We have chosen these two materials as
representatives of an $s$-wave and a (high-temperature) $d$-wave
superconductor, respectively. We have considered a planar multilayered
structure with the respective superconducting layer placed above a
copper substrate. The shielding properties of the two superconductors
have been investigated by defining a screening factor for the atomic
spin relaxation and studying its dependence on the thickness of the
superconducting layer. Despite not having obtained an analytical
solution, we can still affirm that there is an active screening in the
penetration depth layer of the superconductors. When the thickness is
bigger than the penetration depth, the screening saturates. We believe
that these results represent an important step towards the adoption of
neutral atoms as sensitive probes in the study of properties of
different types of superconductors. 

The temperature dependence of the permittivity in $s$-wave and
$d$-wave superconductor follows two different power laws, such
difference has been studied by comparing the spin flip lifetime. When
both niobium and BSCCO are in the superconducting phase, the niobium
gives the longest lifetimes. The comparison has been carried on by
investigating the spin flip lifetime near the superconducting
transition. There is a marked difference between the two
superconductors for $T \leq T_{C}$, which we believe can be observed
experimentally to test the different temperature dependences.  

\acknowledgments
This work was supported by the UK Engineering and Physical Sciences
Research Council (EPSRC), the UK Quantum
Information Processing Interdisciplinary Research Collaboration
(QIP IRC), and the SCALA programme of the European commission.

\appendix
\section{Green function for anisotropic planar multilayers}
\label{sec:green}

In this section, we present our calculation of the Green tensor for an
anisotropic multilayered structure based on the work of Li \textit{et
al.} in Ref.~\cite{Green}. The geometry of the problem is shown in
Fig.~\ref{fig:layerS}b  where the second layer is characterized by a
tensor dielectric permittivity as in Eq.~(\ref{eq.permittivity}). The
Green function will be expressed in terms of an expansion of the
cylindrical vector wave functions adopting the 
cylindric basis
$\{\mathbf{e}_r,\mathbf{e}_{\varphi},\mathbf{e}_z\}$. For an atom
located at $\mathbf{r}'$ in the first layer (vacuum), the Green tensor
can be written as  
\begin{equation}
\label{eq:repG} \bm{G}(\mathbf{r},\mathbf{r}',\omega)=
\bm{G}_{0}(\mathbf{r},\mathbf{r}',\omega)
+\bm{G}_{s}(\mathbf{r},\mathbf{r}',\omega)\,,
\end{equation}
where $\bm{G}_{0}(\mathbf{r},\mathbf{r}',\omega)$ is the unbounded
(bulk) Green tensor representing the contribution of direct waves from
the source at $\mathbf{r}'$ to the point $\mathbf{r}$, while
$\bm{G}_{s}(\mathbf{r},\mathbf{r}',\omega)$ describes the reflection
of waves from the interface between the first and second layer.  

The unbounded Green function can be written as 
\begin{eqnarray}
\lefteqn{
\bm{G}_{0} (\mathbf{r},\mathbf{r}') 
=
-\frac{\delta(\mathbf{r}-\mathbf{r}') }{\omega^{2} \mu_{0} \epsilon_{0}} 
\mathbf{e}_z  \mathbf{e}_z  
+ \frac{i}{4 \pi} 
\int\limits_{0}^{\infty} d \eta 
\sum_{n=0}^{\infty} \frac{2-\delta^{0}_{n}}{ h} 
}
\nonumber 
\\
&&
\left [ \mathbf{M}_{^e _o n }(\pm  h)  \mathbf{M}^{'}_{^e _o n }(\mp  h)
+\left ( \mathbf{N}_{^e _o n  t}(\pm  h) 
+ \mathbf{N}_{^e _o n  z}(\pm  h) \right )
\right.
\nonumber
\\
&& 
\left.
\left ( 
\mathbf{N}^{'}_{^e _o n t}(\mp  h) - \mathbf{N}^{'}_{^e _o n  z}(\mp  h) 
\right ) 
\right ]\,,
\end{eqnarray}
with $h=\sqrt{k^2- \eta^2}$ and $k=\omega/c$.
The scattering dyadic Green function has a form similar to the unbounded Green function and can be formulated as 
\begin{eqnarray}
\lefteqn{\bm{G}_{s} (\mathbf{r},\mathbf{r}')  =
\frac{i}{4 \pi} 
\int\limits_{0}^{\infty} d \eta
\sum_{n=0}^{\infty} \frac{2-\delta_{n}^{0}}{\eta }
}
\\
&&
\left ( \frac{1}{h_{11}}  B_{M 1}^{11}  \mathbf{M}_{^e _o n}( h_{11}) 
\mathbf{M}^{'}_{^e _o n }(h_{11})
\right.
\nonumber
\\
&&
\left. 
-\frac{1}{h_{12}}
\left \{ 
B_{N 2}^{11} 
\left (
 \mathbf{N}_{^e _o n  t}(h_{12}) 
+ \mathbf{N}_{^e _o n  z}( h_{12})
\right )
 \mathbf{N}^{'}_{^e _o n  t}( h_{12}) 
\right. \right.
\nonumber 
\\
&& 
\left. \left.
+ F_{N 2}^{11} 
 \left (\mathbf{N}_{^e _o n t}( h_{12})
+ \mathbf{N}_{^e _o n  z}(h_{12}) \right )\mathbf{N}^{'}_{^e _o n  z}( h_{12}) 
\right \}
\right )\,,
\nonumber
\end{eqnarray}
where in the $l$th layer $h_{lj}=\sqrt{k_{lj}^2 -\eta^2}$
with
\begin{eqnarray}
k_{l1}^{2}&=& \omega^{2} \mu_{0} \epsilon_{tl}\,,
\\
k_{l2}^{2}&=& \eta^{2} \left ( 1-\frac{\epsilon_{tl}}{\epsilon_{zl}}\right )+ \omega^{2} \mu_{0} \epsilon_{tl}\,,
\end{eqnarray}
and $h_{11}=h_{12}=h$.
The cylindrical vector wave functions are defined as 
\begin{eqnarray}
\!\!\!
\label{eq:NM} {\mathbf{M}}_{^e _o n }(h) &=& {\bm{\nabla}} \times
\left[ J_{n}(\eta r) \binom{\cos}{\sin} n \phi\, e^{i h z}
\mathbf{e}_z \right],
\\
\!\!\!{\mathbf{N}}_{^e _o n}(h) &=& \frac{1}{k_{\eta}}{\bm{\nabla}} \times
{\bm{\nabla}} \times \left[ J_{n}(\eta r) \binom{\cos}{\sin} n \phi\,
e^{i h z} \mathbf{e}_z \right ],
\end{eqnarray}
in which $J(\eta r)$ is the Bessel function of the first kind and the
first order, and $k_{\eta}$ is generally defined as
$k_{\eta}= (\eta^{2}+h^{2})^{1/2}$.

The coefficients $B_{M 1}^{11}$, $B_{N 2}^{11}$ and $B_{N 2}^{11}$ are
determined by satisfying the boundary conditions for the Green
function at the interface between layers.  
For the sake of simplicity, we are now going to choose a reference
system such that the coordinate $r$ can be safely set equal to 0 and
the Bessel functions can be expanded for $r \eta \rightarrow 0$ such
that 
$
J_{0} (r \eta)  \approx 1
$ and
$
J_{n}(r \eta) \approx  (r \eta /2)^{n}/n!.
$

The spin flip rate as in Eq.~(\ref{eq:sp}) has been evaluated
throughout this paper for a randomly oriented spin. We consider the
relaxation rate to be the sum of the contribution given by the atomic
spin  parallel or perpendicular to the planar substrate. 
We calculate the double curl of the Green function $ (
\overrightarrow{\bm{\nabla}} \times  \bm{G}(\mathbf{r},\mathbf{r}')
\times \overleftarrow{\bm{\nabla}}' ) _{mn}$ such that $m,n=r,z$, with
the $r$ component coupling to the parallel orientation of the spin and
the $z$ components coupling to the spin perpendicularly oriented.  
The double curl of the 
scattering Green tensor reads
\begin{eqnarray}
\lefteqn{
\overrightarrow{\bm{\nabla}} \times
\bm{G}_{s}(\mathbf{r},\mathbf{r}')  \times
\overleftarrow{\bm{\nabla}}'  
=
}\\
&&
\int\limits_{0}^{\infty} d \eta \, \frac{ ie^{2 i h z}}{4 \pi}
\left [ B_{M 1}^{11} \left ( \frac{\eta^{3}}{h}  - \frac{h \eta}{2} \right )
-B_{N 2}^{11} \frac{\eta k^{2}}{2 h} 
\right ].
\nonumber
\end{eqnarray}
The scattering coefficients for a general argument $h_{f}$ are given by
\begin{eqnarray}
\lefteqn{
B_{M,N}^{11} = \pm \frac{e^{-2 i h_{1} d }}{1+ R_{1}^{H,V} R_{2}^{H,V}
e^{i 2 h_{2} d }} 
}
\nonumber
\\
&& \left [
R_{1}^{H,V} + R_{2}^{H,V} e^{2 i h_{2} d } \right]\,,
\end{eqnarray}
with 
\begin{eqnarray}
R_{f}^{H} &=& \frac{h_{f+1}-h_{f}}{h_{f+1}+h_{f}} \,,
\\
\nonumber \\
R_{f}^{V}&=& \frac{
\frac{h_{f} [ (\omega_{1}-\omega_{2}) h_{f+1}^{2} +\omega_{2}
k_{f+1}^{2} ]   }{h_{f+1} [(\omega_{1} -\omega_{2}) h_{f}^{2}
+\omega_{2} k_{f}^{2}]} -1 
}
{\frac{h_{f} [ (\omega_{1}-\omega_{2}) h_{f+1}^{2} +\omega_{2}
k_{f+1}^{2} ]   }{h_{f+1} [(\omega_{1} -\omega_{2}) h_{f}^{2}
+\omega_{2} k_{f}^{2}]} +1 
} \,,
\end{eqnarray}
where  $d$ is the height of the second layer, and $\omega_{1}$ and
$\omega_{2}$ are the weighting coefficient that in our case equal to
1. To obtain the specific scattering coefficients, $h_{f}$ has to be
substituted by $h_{f1}$ or~$h_{f2}$.

\end{document}